\documentclass[review,authoryear,3p,12pt]{elsarticle}

\usepackage{amssymb,amsmath}
\usepackage{color}

\usepackage[utf8]{inputenc}

\usepackage{natbib}
\usepackage{graphicx}
\usepackage{longtable}
\usepackage{caption}
\usepackage{lineno}

\biboptions{authoryear,comma,round}

\usepackage{color}
\usepackage[normalem]{ulem}    % For strikethrough, text
\usepackage{cancel}    	       % For strikethrough, math mode
\definecolor{add}{rgb}{0.0,0.0,0.0}
\definecolor{del}{rgb}{0.3,0.5,1.0}
    
\newcommand{\add}[1]{{\color{add} {#1}}} % track changes: add new text

\begin{document}

\begin{frontmatter}

\title{Can a fractionally crystallized magma ocean explain the thermo-chemical evolution of Mars?}
%\title{Consequences of magma ocean cumulate overturn on mantle dynamics: thermo-chemical evolution of Mars}

\author[1,2]{A.-C. Plesa}\author[1,3]{N. Tosi}\author[1]{D. Breuer}

\address[1]{Institute of Planetary Research, German Aerospace Center (DLR), Rutherfordstr. 2, 12489 Berlin, Germany (ana.plesa@dlr.de)}
\address[2]{Institute of Planetology, University of M\"unster, Schlossplatz 2, 48149 M\"unster, Germany}
\address[3]{Department of Planetary Geodesy, Technische Universit\"at Berlin, Strasse des 17. Juni 135 10623 Berlin, Germany}

\begin{abstract}
The impact heat accumulated during the late stage of planetary accretion can melt a significant part or even the entire mantle of a terrestrial body, giving rise to a global magma ocean. The subsequent cooling of the interior causes the magma ocean to freeze rapidly from the core-mantle boundary (CMB) to the surface due to the steeper slope of the mantle adiabat compared to the slope of the solidus. 

Assuming fractional crystallization of the magma ocean, dense cumulates are produced close to the surface, largely due to iron enrichment in the evolving magma ocean liquid \citep{elkinstanton03}.  A gravitationally unstable mantle thus forms, which is prone to overturn. We investigate the cumulate overturn and its influence on the thermal evolution of Mars using mantle convection simulations in 2D cylindrical geometry. We present a suite of simulations using different initial conditions and a strongly temperature-dependent viscosity. We assume that all radiogenic heat sources have been enriched during the freezing-phase of the magma ocean in the uppermost $50$ km and that the initial steam-atmosphere created by the degassing of the freezing magma ocean was rapidly lost, implying that the surface temperature is set to present-day values. In this case, a stagnant lid forms rapidly on top of the convective interior preventing the uppermost dense cumulates to sink, even  when allowing for a plastic yielding mechanism. Below this dense stagnant lid, the mantle chemical gradient settles to a stable configuration. The convection pattern is dominated by small-scale structures, which are difficult to reconcile with the large-scale volcanic features observed over Mars' surface and partial melting ceases in less than $900$ Ma. Assuming that the stagnant lid can break because of additional mechanisms and allowing the uppermost dense layer to overturn, a stable density gradient is obtained, with the densest material and the entire amount of heat sources lying above the CMB. This stratification leads to a strong overheating of the lowermost mantle, whose temperature increases to values that exceed the liquidus. The iron-rich melt would most likely remain trapped in the lower part of the mantle. The upper mantle in that scenario cools rapidly and only shows partial melting during the first billion year of evolution. Therefore a fractionated global and deep magma ocean is difficult to reconcile with observations. Different scenarios assuming, for instance, a hemispherical or shallow magma ocean, or a crystallization sequence resulting in a lower density gradient than that implied by pure fractional crystallization will have to be considered.

\end{abstract}


\end{frontmatter}

%\linenumbers

%%                Introduction
%%%%%%%%%%%%%%%%%%%%%%%%%%%%%%%%%%%%%%%%%%%%%%%%%%%%%%%%%%%%%%%%%%%%%%%%%%%%%%%%%%%%%%%%%%%%%%%
\section{Introduction}
\label{Introduction}

Geochemical analyses of the so-called SNC meteorites suggest the existence of three to four separate and isotopically distinct reservoirs in the martian mantle, which have been preserved over the entire planetary evolution \citep[e.g.,][]{jagoutz91, papike09}. Two of these reservoirs are depleted in incompatible elements and are most likely situated in the mantle, whereas the third one is enriched and could be located either in the crust or in the mantle \citep{foley05}. Until present, most of the dynamical simulations of Mars' mantle convection did not account for the formation or preservation of such reservoirs, and focused mainly on explaining, for instance, the crustal dichotomy or the Tharsis bulge \citep[e.g.,][]{breuer98, harder96, schumacher06, keller09, sramek12}. Scenarios that consider the early formation of chemical heterogeneities in the Martian mantle assume either a global magma ocean \citep[]{elkinstanton03, elkinstanton05b, debaille09}, or at least substantial partial melting in the earliest phase of the planet's evolution \citep[]{schott01, ogawa11, plesa13a}. It is suggested that early in the evolution of Mars, the large amount of primordial heat due to accretion, core formation, and possibly short lived radioactive elements can give rise to a magma ocean as a consequence of significant or perhaps even complete melting of the mantle \citep[e.g.,][]{breuer07}. This assumption is confirmed by studies on short-lived radionuclides such as $^{182}$Hf suggesting that the separation between silicates and iron occurred in the first million years after accretion \citep[e.g.,][]{kleine02}. Furthermore, estimates of the timescale of the process of core formation suggest that, to achieve rapid separation, both silicates and metals need to be fluid, at least in the upper part of mantle \citep[e.g.,][]{stevenson90}.

In general, the magma ocean freezes from the bottom (i.e. from the CMB, if it comprises the entire mantle) to the surface because of the steeper slope of the mantle adiabat compared to the slope of the solidus \citep[e.g.,][]{solomatov00}. \citet{solomatov00} discusses two main freezing mechanisms: equilibrium and fractional crystallization, depending on the size and the settling velocity of crystals that form upon solidification. In the case of equilibrium crystallization, the crystal size and the settling velocities are so small that freezing takes place before crystal-melt separation occurs. The magma ocean solidifies without differentiating and a chemically homogeneous mantle is formed. In the case of fractional crystallization, instead, denser cumulates are formed while the crystallization of the magma ocean proceeds from the bottom, mainly because of iron enrichment in the evolving magma ocean liquid \citep{elkinstanton03}. The consequence is the formation of an unstably stratified density profile. Whether equilibrium or fractional crystallization in a magma ocean is the dominant process also depends on the time scale of freezing and may change with time. A large degree of melting of the mantle implies a very low viscosity of the magma ocean. The viscosity of a melt/crystal mixture increases abruptly near a critical crystal fraction as suggested by theoretical and experimental studies of concentrated suspensions \citep[e.g.,][]{mooney51, roscoe52, campbell90} and by experiments with partial melts \citep[e.g.,][]{arzi78, lejeune95}. This sudden increase in viscosity is defined by the so-called “rheological transition”,\add{ which} depends on the crystal size distribution, the crystal shape\add{ and orientation}, and other factors. \add{The critical crystal fraction for this transition varies in the literature. \citet{saar01}, using numerical models, found that for plagioclase crystals, the critical crystal fraction that allows a crystal network to form lies between $8$ and $20$ \%. Other studies place this transition at $30$\% to $50$\% depending on the basalt composition \citep{philpotts96}. In \citet{solomatov00}, the rheological transition is assumed to take place at $\sim 60$ \%.} \citet{solomatov00} argues for instance that freezing of the Earth's lower \add{mantle} lasts only a few hundred years, until the crystal fraction becomes larger than the critical value marking the rheological transition. For larger crystal fractions, the time scale for further crystallization slows down by several orders of magnitudes ($\sim 10$ to $100$ Ma) as the convection speed and cooling efficiency of the partially molten mantle is significantly reduced. Fractional crystallization may become dominant at this stage. \add{If the crystals do not remain in suspension, a crystal network forms instead and efficiently compacts, ultimately resulting in a fractionation between crystals and melt}. Therefore, depending on how the magma ocean freezes, i.e. via fractional or equilibrium crystallization,  either a chemically homogeneous or stratified mantle is obtained and different evolutionary paths can be expected. So far, two end-member scenarios have been suggested in the literature for Mars: 

1) In the first scenario, it is assumed that\add{ the early martian mantle is homogeneous apart from an upper depleted layer  \citep{schott01, ogawa11}.  Such a structure can evolve if Mars did not experience a deep magma ocean but a large amount of melting causing secondary differentiation early in its evolution. Alternatively, if a deep magma ocean existed, the largest part of the mantle, from the CMB up to a certain depth, freezes rapidly by equilibrium crystallization and only melt from a remaining shallow layer rises via porous flow due to compaction of the silicate matrix or through cracks and channels.} This early melt produces a primordial crust and marks the start of the differentiation process. As the mantle residue is in general less dense than the primordial mantle, chemically distinct reservoirs can form depending on the density contrast between depleted and primordial mantle \citep{plesa13a}. Typically, a depleted and compositionally buoyant upper mantle can develop, which overlies an undepleted and compositionally denser lower mantle \citep{schott01, ogawa11}. These two layers convect separately with the lower one that tends to become thinner with time because of erosion, while its composition remains almost primordial. However, whether or not this scenario can actually explain the isotopic characteristics of the martian meteorites is unclear and, to the best of our knowledge, it is a problem that still needs to be investigated. 

2) In the second scenario, the mantle freezes by fractional crystallization resulting in a stratification with Fe-rich cumulates close to the surface and Mg-rich cumulates at the CMB, a configuration that is gravitationally unstable. Being highly incompatible, radioactive elements are enriched in the evolving residual liquid of the magma ocean.  Therefore, a substantial amount of radioactive heat-sources is concentrated in the uppermost layer of Mars' mantle \citep{elkinstanton05a}. The unstable layering results then in an global mantle overturn, which has been estimated to take place within $\sim 10-100$ Ma \citep{elkinstanton03,elkinstanton05a,elkinstanton08,debaille09}.  As a consequence of the overturn, the dense surface material sinks to the core-mantle boundary, while the light material from the CMB rises to the surface, thus leading to a stable, chemically layered mantle. It is further assumed that, during the overturn, rising upwellings can melt by adiabatic decompression and produce an early crust. The residual mantle of this process has been associated with the source region of the nakhlites \citep{debaille09}. The chemical layering following the overturn suppresses subsequent thermally driven convection and early-formed reservoirs are preserved throughout the planet's history. In this way, distinct reservoirs can be maintained stable over the rest of the thermal evolution of Mars. 

An important feature of the mantle overturn is also the subduction of the dense surface material. The first models with which this process was investigated \citep{elkinstanton03,elkinstanton05b,elkinstanton05a} assumed a small viscosity contrast between the interior and the surface, implicitly allowing for the mobilization of the surface layers. A small viscosity contrast can be the consequence of high surface temperatures maintained during the solidification phase of the magma ocean. In fact, a large amount of volatiles is expected to be outgassed, thereby forming a steam atmosphere able to keep the surface temperature even above the solidus of silicates \citep{abe93, elkinstanton05a, elkinstanton08}. However, whether and how long a thick proto-atmosphere can be preserved is unclear since atmospheric loss processes are likely to be enhanced in the early stage of planetary evolution due to a stronger solar activity. \citet{lebrun13}\add{, using a convective-radiative atmospheric model in equilibrium with the magma ocean, showed that a thick proto-atmosphere could be lost in about $10$ Ma. A recent study based on a 1D hydrodynamic upper atmosphere model taking into account the extreme XUV conditions of the young Sun suggests that a thick atmosphere produced by outgassing during magma ocean crystallization can be lost in $0.14$ to $14$ Ma }\citep{erkaev13}\add{. The time needed to overturn the densest cumulates has been computed assuming a constant viscosity of $10^{20}$ Pa\, s and was found to be a fraction of a million year }\citep{elkinstanton03}\add{, thus comparable with the time predicted for the loss of the initial atmosphere. However, even assuming that a high surface temperature ($900^{\circ}$C according to }\citet{elkinstanton05b}\add{) can be maintained for longer than $10$ Ma, when using an activation energy of $300$ kJ/mol, as expected for an olivine-dominated rheology, the resulting surface viscosity would be about $8$ orders of magnitude higher than the bulk viscosity of the interior, which still represent a major obstacle for the mobilization of the uppermost cumulates.}

Another possibility to obtain surface mobilization, even in the presence of low surface temperatures, is by lithospheric failure. Only recently it has been shown that rising upwellings in the early stage of the overturn can localize surface deformation, inducing lithospheric-scale yielding and consequent surface mobilization \citep{debaille09, tosi13}. These models, however, used for simplicity either a linear unstable density profile \citep{tosi13}, or a density profile in which phase transitions were neglected \citep{debaille09}. Furthermore, an approximation of the viscosity law, the so-called Frank-Kamenetskii approximation (F-K), which can result in an underestimation of shallow viscosity gradients, is commonly used \citep{debaille09}. 

\add{In the present work, in contrast to previously published results that mainly focused on the overturn itself, we further investigate the consequences of a magma ocean cumulate overturn on the subsequent thermo-chemical evolution of Mars.} We will show  that considering a density profile that accounts for phase transitions, as well as a more appropriate strongly temperature-dependent rheology and a plastic deformation mechanism, the overturn tends to be confined below a stagnant lid and does not lead to a mobilization of the surface layers. \add{Even if a whole mantle overturn takes place, assuming additional mechanisms discussed below, the density gradient obtained after the overturn suppresses convective heat transport. In both cases, i.e. whole mantle overturn and overturn below the stagnant lid, the subsequent thermo-chemical evolution is hard to reconcile with observations.} The findings of the present study demonstrates the difficulty to find plausible scenarios to fit results from geochemical and isotope analyses with those of geodynamical modelling.

%We find that if using a non-linear rheology with realistic parameters the cumulate overturn will not take place as modelled in previous studies and a high density layer will remain at the surface. If the upper layer breaks by additional mechanisms, then the achieved density gradient after the overturn suppresses the thermal convection. This is an important aspect since the volcanic history of Mars \citep{neukum04} hints at an active interior up to the recent past.

%%%%%%%%%%%%%%%%%%%%%%%%%%%%%%%%%%%%%%%%%%%%%%%%%%%%%%%%%%%%%%%%%%%%%%%%%%%%%%%%%%%%%%%%%%%%%%%
%%				Model
%%%%%%%%%%%%%%%%%%%%%%%%%%%%%%%%%%%%%%%%%%%%%%%%%%%%%%%%%%%%%%%%%%%%%%%%%%%%%%%%%%%%%%%%%%%%%%%
\section{Model and Methods}\label{Model and Methods}
\subsection{Physical Model}\label{Physical Model}
We assume that the amount of heat generated by accretion and core formation processes has been large enough during the early planetary evolution to melt the whole mantle, creating a global magma ocean. Subsequent cooling and fractional crystallization of the magma ocean results in a stratified and gravitationally unstable mantle. We assume that the initial steam atmosphere created during the magma ocean crystallization phase is entirely lost at the time when the magma ocean has fully crystallized (\citet{lebrun13, erkaev13}), such that the surface temperature is set to a present day value of $250$ K. 

\add{Radiogenic elements, being highly incompatible, will preferentially partition in the evolving liquid phase during the fractional crystallization of the magma ocean. Following the model by \citet{elkinstanton05b}, we set the entire amount of heat producing elements in the uppermost $50$ km.}

\begin{figure}[!ht]
 \begin{center}
    \includegraphics[width=\textwidth,angle=0]{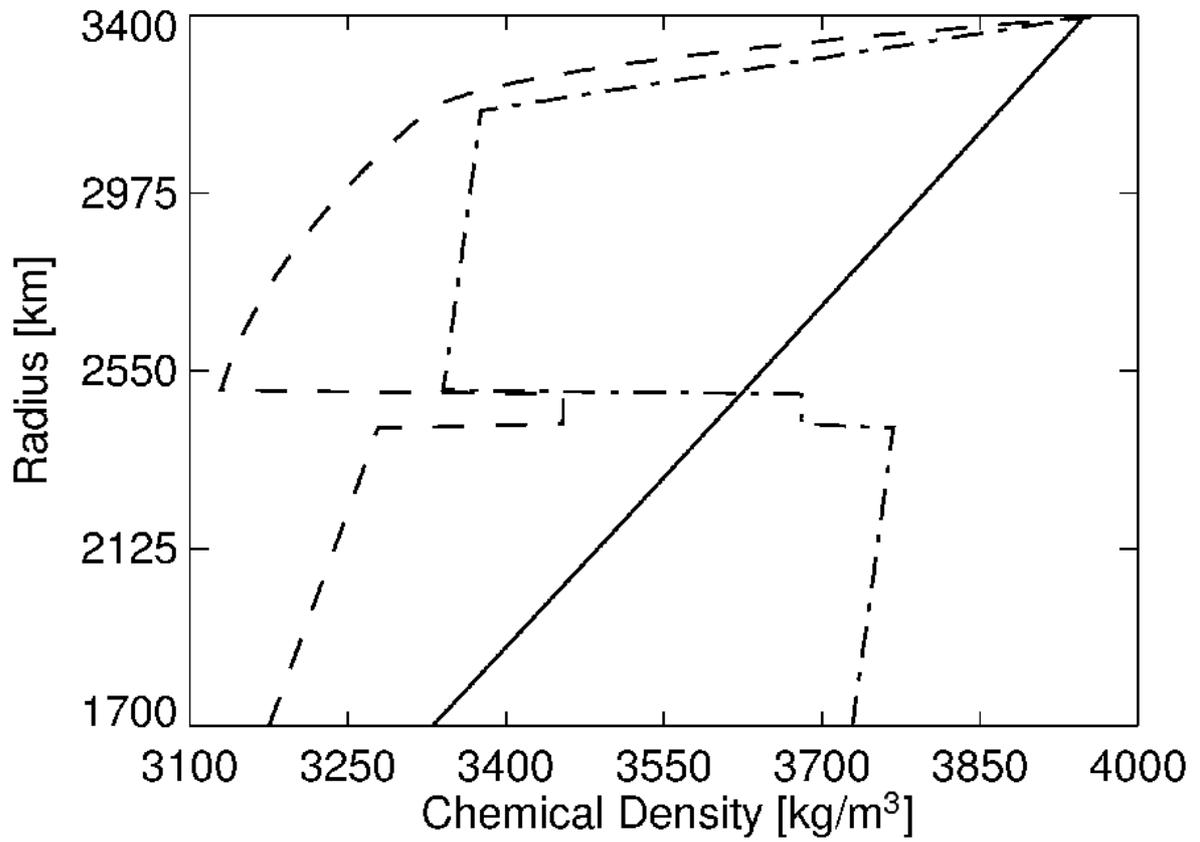}
    \caption{Initial density profiles. The full line refers to a linear profile with the same global density jump prescribed for the dashed-dotted profile. The dashed line refers to the density 
     profile calculated by \citet{elkinstanton05b}. The dashed-dotted line refers a density profile calculated using densities from \citet{elkinstanton03} and accounting for phase transitions.}
     \label{fig_all_profiles}
 \end{center}
\end{figure}

The density profile established after freezing of the magma ocean is a crucial parameter, since it can dominate the entire dynamic and even suppress thermal buoyancy. However, such a profile is poorly constrained and varies in the literature from a linear function of the mantle depth \citep{zaranek04} to a more complex one, strongly depending on the mantle mineralogy \citep{elkinstanton03}. In order to test the effects of the initial density profile and to compare our results with previous models, we compare in the following three different profiles (Figure \ref{fig_all_profiles}). 

We run one simulation assuming an unstable linear profile. For this we use the density contrast from \citet{elkinstanton05b} calculated at a reference temperature of $1^{\circ}$ C. Therefore the linear profile has the same buoyancy ratio as the third profile presented below.

\begin{figure}[!ht]
 \begin{center}
    \includegraphics[width=\textwidth,angle=0]{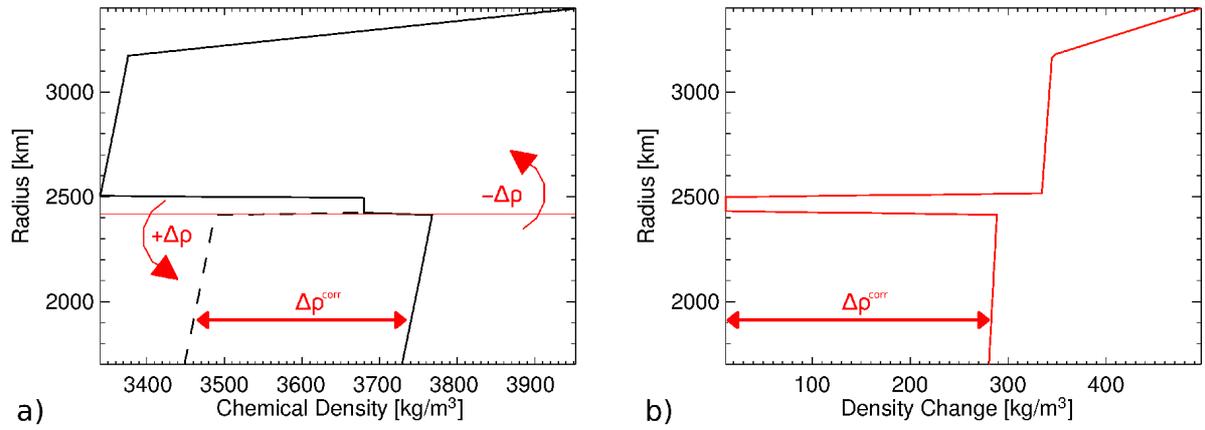}
    \caption{Density profile calculation. a) Full line represents the density profile for Mars-like parameters calculated using the densities from \citet{elkinstanton03} and b) the corresponding density change upon mineralogical phase change calculated after \citet{elkinstanton03}. Dashed line in a) shows the density in the lower part of the mantle calculated under the assumption that whole-mantle overturn will take place. This assumption has been made in previous studies in which the actual lower mantle density was decreased by $\Delta \rho^{corr}$ 
    \citep[][]{debaille09,elkinstanton05b}.}\label{fig_magmaocean_profile}
 \end{center}
\end{figure}

A second run is performed using the initial density profile from \citet{elkinstanton05b} that also considers the formation of a dense garnet layer in the mid-mantle. The models presented in \citet{elkinstanton03} and in subsequent studies \citep{elkinstanton05a,elkinstanton05b,debaille09} assume a simplified approach appropriate to simulate the overturn but not the subsequent thermal evolution: major mantle phase transitions are only indirectly considered by shifting the lower mantle material above the phase transition depth, adjusting its density profile accordingly and setting the initial density value in the lower mantle to the recalculated profile. Therefore, although majorite and $\gamma$-olivine are actually the minerals in the lower mantle, the density profile is recalculated using the corresponding volume fraction and the density values of olivine, pyroxene and garnet -- main constituents of the upper mantle (see dashed line in Figure \ref{fig_magmaocean_profile}a). Furthermore, the dashed profile in Figure \ref{fig_all_profiles} accounts for the material being at the solidus temperature -- thus, the density (dashed line in Figure \ref{fig_magmaocean_profile}a) is further reduced in contrast to the mantle material at reference conditions.      

In a third simulation, we compute the initial density distribution using the reference density values from \citet{elkinstanton03} and the mantle mineralogy, i.e. magnesium number and volume percent of each phase (olivine, garnet, pyroxene, majorite and $\gamma$-olivine) from \citet{elkinstanton05b}. The resulting profile is shown in Figure \ref{fig_magmaocean_profile}a. Since we are also interested in the evolution after the magma ocean crystallization and overturn, we use densities at $1$ atm and $1^{\circ}$ C rather than $1$ atm and solidus temperature, which would be valid only in the overturn phase. Additionally, we account in our models for the transition from garnet and pyroxene to majorite and olivine to $\gamma$-olivine. As in the model of \citet{elkinstanton05b}, we assume for simplicity that all phase transitions occur at a pressure of $\sim 13$ GPa equivalent to the depth of $\sim 1000$ km in Mars. We calculate the density difference of the mantle material according to the magnesium number and the volume percent of each phase (Figure \ref{fig_magmaocean_profile}b). Therefore, when material of the shallow regions of the mantle sinks to the core-mantle boundary, garnet, pyroxene and olivine increase their density as a consequence of the transition to majorite and $\gamma$-olivine. For material from the deep regions containing $\gamma$-olivine and majorite and rising to shallower depths, the density is reduced according to the phase transition. Although the transition from pyroxene and garnet to majorite is slightly endothermic, on average, the transition of all phases is exothermic (\citet{elkinstanton03}). The exothermic behavior is taken into account in such a way that the phase transition in cold downwellings takes place at a shallower depth compared to hot upwellings.

\subsection{Mathematical Formulation}\label{Mathematical Formulation}
We model the thermo-chemical convection of Mars' mantle by solving the conservation equations of mass, momentum, energy and composition. These are scaled as usual using the thickness of the mantle $D$ as length scale, the thermal diffusivity $\kappa$ as time scale, the temperature drop $\Delta T$ and the maximum chemical density contrast $\Delta \rho$ across the mantle as temperature and compositional scales. Using the extended Boussinesq approximation (EBA), a Newtonian rheology and an infinite Prandtl number, as appropriate for highly viscous media with negligible inertia, the non-dimensional equations of thermo-chemical convection read:
\begin{eqnarray}
\nabla \cdot \vec{u} &=& 0\label{eq_mantle_convection1}\\
\nabla \cdot \left[\eta(\nabla \vec{u} + (\nabla \vec{u})^T)\right] + (Ra T - Ra_C C) \vec{e}_r - \nabla p &=& 0\label{eq_mantle_convection2}\\
\frac{\partial T}{\partial t} + \vec{u} \cdot \nabla T - Di(T+T_{surf})u_r - \nabla ^2 T - \frac{Di}{Ra}\Phi - H &=& 0\label{eq_mantle_convection3}\\
\frac{\partial C}{\partial t} + \vec{u}\cdot \nabla C &=& 0\label{eq_mantle_convection4}
\end{eqnarray}
where $\vec{u}$ is the velocity vector, $p$ is the dynamic pressure, $T$ is the temperature, $t$ is the time, $\eta$ is the viscosity, $u_r$ is the radial velocity and $\vec{e}_r$ is the radial unit vector. $Ra$ and $Ra_C$ are the thermal and compositional Rayleigh number respectively, $H$ is the internal heating rate, which is given by the ratio of $Ra_Q$ and $Ra$, where $Ra_Q$ is the Rayleigh number for internal heat sources (see Table \ref{tab_parameters}). The three Rayleigh numbers read:
\begin{equation}\label{eq_Rayleigh}
Ra = \frac{\rho g \alpha \Delta T D^3}{\eta \kappa}\hspace{5mm} \text{, } \hspace{5mm} Ra_C = \frac{\Delta \rho g D^3}{\eta \kappa}\hspace{5mm} \text{, } \hspace{5mm} Ra_Q = \frac{\rho^2 g \alpha H D^5}{\eta k \kappa}
\end{equation}
In all simulations we use a fixed surface temperature, a cooling boundary condition at the core-mantle boundary, and decaying radioactive heat sources. Similarly to the model of 
\citet{morschhauser11}, the concentration of long-lived elements is taken from \citet{waenke94}. The viscous dissipation is defined as follows:
\begin{equation}
\Phi = 2 \eta \dot{\varepsilon}^2
\end{equation}
The dissipation number, $Di$, depends on the thermal expansivity $\alpha$, the gravity acceleration $g$, the mantle thickness $D$ and the heat capacity of the mantle $c_p$:
\begin{equation}
Di=\frac{\alpha g D}{c_p}
\end{equation}
All variables and parameters used in equations (\ref{eq_mantle_convection1})-(\ref{eq_Rayleigh}) are listed in Table \ref{tab_parameters}.

The viscosity is calculated according to the Arrhenius law for diffusion creep \citep{karato86}. The non-dimensional formulation of the Arrhenius viscosity law for temperature dependent viscosity reads \citep[e.g.][]{roberts06}:
\begin{equation}\label{eq_viscosity}
\eta(T) = \exp\Big(\frac{E}{T+T_{surf}} - \frac{E}{T_{ref}+T_{surf}}\Big),
\end{equation}
where $E$ is the activation energy, $T_{surf}$ the surface temperature and $T_{ref}$ the reference temperature at which a reference viscosity is attained (see Table 1).

In some cases, we use a linearization of equation (\ref{eq_viscosity}), the so-called Frank-Kamenetskii (F-K) approximation. For temperature dependent viscosity, this can be written as:
\begin{equation}\label{eq_viscosity_FK}
\eta(T) = \exp\Big(\gamma(T - T_{ref})\Big),
\end{equation}
where $\gamma$ is the F-K parameter \citep{solomatov97}, which in our simulations is set to $10^5$ as in \citet{debaille09}.

The reference viscosity is set to $10^{22}$ Pa s at a reference temperature of $1600$ K which results in a lower mantle viscosity of $\sim 10^{18} - 10^{19}$ Pa s. \add{It should be noted that such reference viscosity, typical of dry materials, is likely too high for Mars' mantle, which is expected to be characterized by a wet rheology \citep{grott13} and hence by a lower viscosity. Nevertheless, this choice only affects the timescale of mantle overturn (Sections \ref{subsect_density_profiles} and \ref{subsect_viscosity}) and does not impact our main conclusions.} A mechanism of pseudo-plastic yielding is used to allow the stagnant lid to self-consistently fail if the convective stresses are high enough to overcome a depth dependent yield stress $\sigma_{y}$. For this, the effective viscosity reads:
\begin{equation}\label{eq_viscosity_PT}
\frac{1}{\eta_{eff}} = \frac{1}{\eta(T)} + \frac{1}{\displaystyle\frac{\sigma_{y}}{2\dot{\varepsilon}}}, 
\end{equation}
where $\sigma_y = \sigma_0 + z\partial \sigma /\partial z$, $z$ is the depth, and $\dot{\varepsilon}$ is the second invariant of the strain rate tensor. For the depth-dependent yield stress, 
we use a surface value $\sigma_0=10^8$ Pa and a gradient $\partial \sigma / \partial z = 160$ Pa/m \citep{nakagawa05}.

\subsection{Numerical Model}\label{Numerical Model}
To solve equations (\ref{eq_mantle_convection1})-(\ref{eq_mantle_convection4}) we used the finite volume code Gaia \citep{huettig08a}. All simulations were run in a 2D-cylindrical geometry on a structured grid consisting of $170$ equally spaced shells in the radial direction and $880$ points in the lateral direction (corresponding to a lateral resolution of $10$ km in the mid of the mantle) resulting in a total of $\sim 1.5 \times 10^5$ grid points. The cylindrical domain was also rescaled to match the inner- to outer-surface ratio of a three-dimensional spherical
shell \citep{vanKeken01}.

Equations (\ref{eq_mantle_convection1})-(\ref{eq_mantle_convection3}) are solved on a fixed mesh. To solve the advection equation (\ref{eq_mantle_convection4}), we use a particle-in-cell method (PIC) \citep[e.g.,][]{vanKeken97, tackley03}. This minimizes spurious effects due to numerical diffusion, which typically arise when using grid-based methods. In all tests, we assigned $20$ particles to each cell, resulting in a total of $\sim 3 \times 10^6$ particles. Our PIC implementation has been tested using various benchmark runs of increasing complexity in both 2D cylindrical and 3D spherical geometry \citep{plesa13}. In our simulations each particle has its own density ($C$) value. Furthermore, when accounting for an exothermic phase transition, the particles are additionally characterized by a phase parameter and a density difference ($\Delta C$). At each time-step and for each particle, the following phase function is computed \citep{christensen85}:
\begin{equation}\label{eq_phases}
\Gamma_0 = \frac{1}{2}\Big(1 + \tanh\Big(\frac{z - z_0(T)}{w}\Big)\Big) 
\end{equation}
where $z_0(T) = z_0 + \gamma_0(T - T_0)$ is the temperature-dependent depth of the phase boundary, $\gamma_0$, $z_0$, $T_0$, and $w$ are the Clapeyron slope, reference depth, reference temperature,
and width of the phase transition, respectively. When a particle crosses the phase boundary calculated from equation (\ref{eq_phases}), it is assigned an appropriate density jump.
%the density difference is applied to the particle's density (either added, if the phase parameter is $0$ or subtracted, if the phase parameter is $1$) and the phase parameter is then updated depending on the particle location (if the particle was previously in the lower mantle with the phase parameter set to $1$, this parameter will be set to $0$ if the particle resides now in the upper part of the mantle and vice versa). 
Note that we neglected the release and consumption of the latent heat associated with the phase change. For an exothermic transition like the one assumed here, this results in an 
slight overestimation of its enhancing effect on mantle flow.

%%%%%%%%%%%%%%%%%%%%%%%%%%%%%%%%%%%%%%%%%%%%%%%%%%%%%%%%%%%%%%%%%%%%%%%%%%%%%%%%%%%%%%%%%%%%%%%
%%				Results
%%%%%%%%%%%%%%%%%%%%%%%%%%%%%%%%%%%%%%%%%%%%%%%%%%%%%%%%%%%%%%%%%%%%%%%%%%%%%%%%%%%%%%%%%%%%%%%
\section{Results}
\label{Results}

We apply the thermo-chemical mantle convection model to simulate magma ocean cumulate overturn assuming three different density profiles (Figure \ref{fig_all_profiles}). If not otherwise specified, the viscosity is calculated using an activation energy of $300$ kJ/mol, as appropriate for dry olivine \citep{karato86}, and applying the Arrhenius law (equation (\ref{eq_viscosity})). \add{Despite the presence of a high concentration of heat producing elements near the surface,} this causes the formation of a stagnant lid, which can be mobilized when considering a plastic rheology (equation (\ref{eq_viscosity_PT})). The initial temperature profile is the same in all tests and follows the mantle solidus, apart for a thin upper thermal boundary layer with a thickness of $20$ km (black 
line in Figure \ref{fig_density_profiles}a). 

\subsection{Effects of the Initial Chemical Density Profile}\label{subsect_density_profiles}

\begin{figure}[!ht]
 \begin{center}
    \includegraphics[width=\textwidth,angle=0]{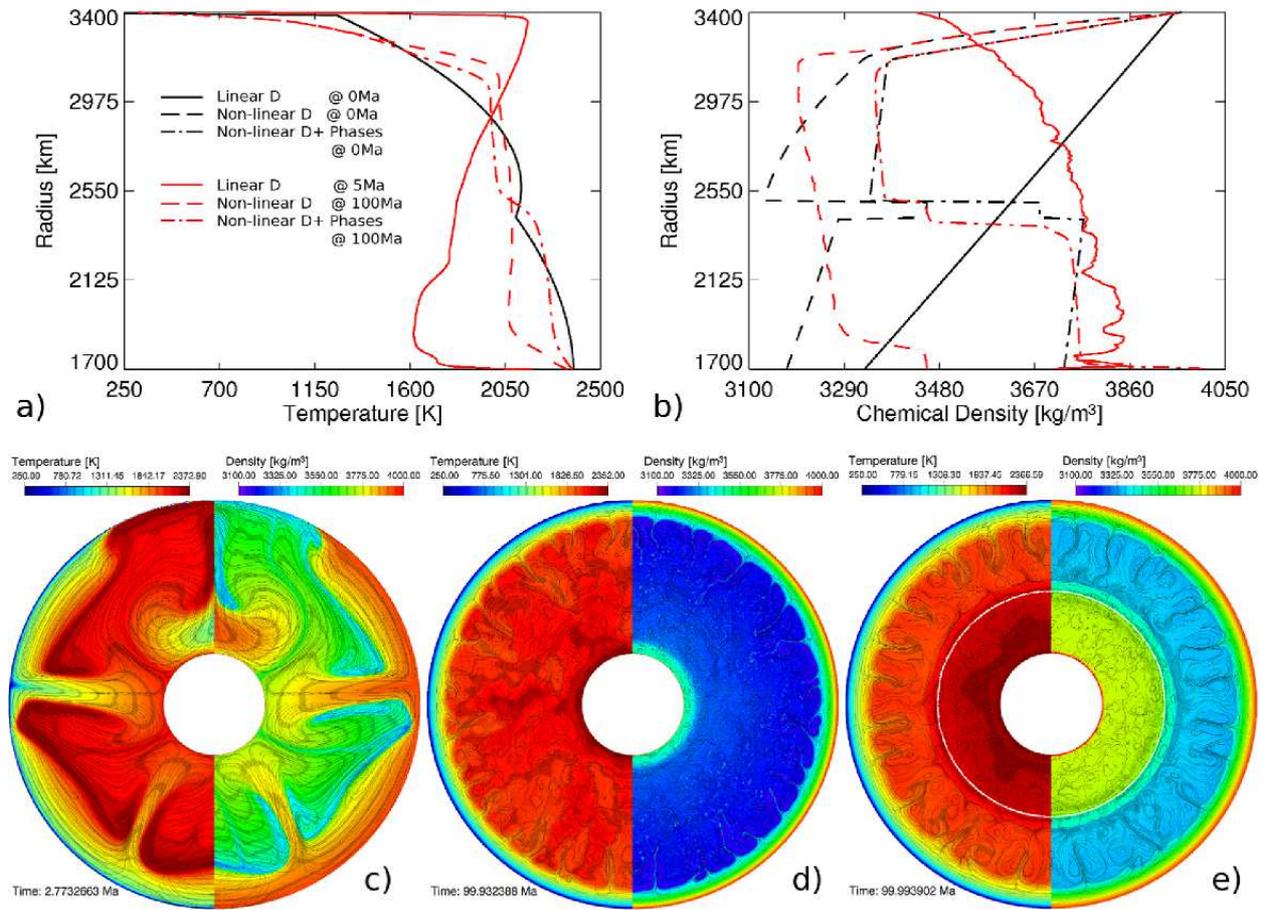}
    \caption{Mantle overturn resulting from the three initial density profiles shown in Figure \ref{fig_all_profiles}. a) Temperature and b) density profiles before (black) and after (red) the     overturn at $100$ Ma. Temperature and chemical density distribution resulting from c) the linear profile, d) the profile after \citet{elkinstanton05b}, and e) the modified profile accounting for phase transitions. Only in the first case a whole-mantle overturn takes place. In the other two cases the overturn occurs below the stagnant lid.} 
    \label{fig_density_profiles}
 \end{center}
\end{figure}

We investigate first the influence of the initial chemical density profile on the cumulate overturn of an unstably stratified mantle using a strongly temperature-dependent viscosity and taking into account plastic behaviour. As shown in Figure \ref{fig_density_profiles}, we consider 1) a linear profile, 2) the profile from \citet{elkinstanton05a}, and 3) a profile after \citet{elkinstanton03} calculated at reference conditions and accounting for mineralogical phase transitions.

A simple linear profile has been widely used in the literature \citep{hansen00,hansen95,zaranek04}. In contrast to previous studies in which simulations are started from a stable configuration, i.e. after the magma ocean overturn, here we start with an unstable profile with the densest material located at the top. Due to the temperature dependence of the viscosity, dense material lying initially just beneath the stagnant lid sinks into the deep mantle while lighter material rises to the surface. The convective stresses created by both the light material which rises to shallower depths and the return flow due to the sinking of heavy cumulates are high enough to overcome the imposed yield stress. The lithosphere experiences plastic yielding and the initially unstable chemical gradient evolves into a stable configuration. Figure \ref{fig_density_profiles}a) and b) (full lines) show the temperature and chemical density profiles after $5$ Ma. At this time, the overturn is nearly completed. During such whole-mantle overturn, relatively long-wavelength convection, involving all mantle layers is observed. Afterwards, small scale convection patterns prevail and may persist for some time until eventually a perfect layering is reached. Note that the thermal buoyancy is not strong enough to overcome the stable chemical gradient during the $4.5$ Ga. All radioactive heat sources are located at the core-mantle boundary layer and heat significantly the lowermost mantle.

The second profile corresponds to the fractional crystallization sequence, including a dense garnet layer in the mid-mantle, but neglects the effect of a mineralogical phase transition and is calculated at the solidus temperature as in \citet{elkinstanton05b}. Although this density profile is unstable as the linear density profile, whole-mantle overturn does not occur within $100$ Ma and excludes the upper dense layer. \add{Using the same yield stress distribution,} surface mobilization is not observed in contrast to the linear density profile. The cause for this is that the densest material is trapped in the stagnant lid below which the chemical gradient is not large enough to generate the necessary convective stresses for breaking the lithosphere. The overturn takes place only below the lid, while the entire amount of heat sources and the densest materials remain at the top. We have also tested the robustness of this finding running additional calculations and assuming a thinner initial thermal boundary layer. As the stagnant lid grows rapidly, similar results have been obtained showing that dense and enriched materials remain in the stagnant lid. Upon  overturn, the dense garnet layer located in the mid-mantle sinks to the core-mantle boundary.

\begin{figure}[!ht]
 \begin{center}
    \includegraphics[width=\textwidth,angle=0]{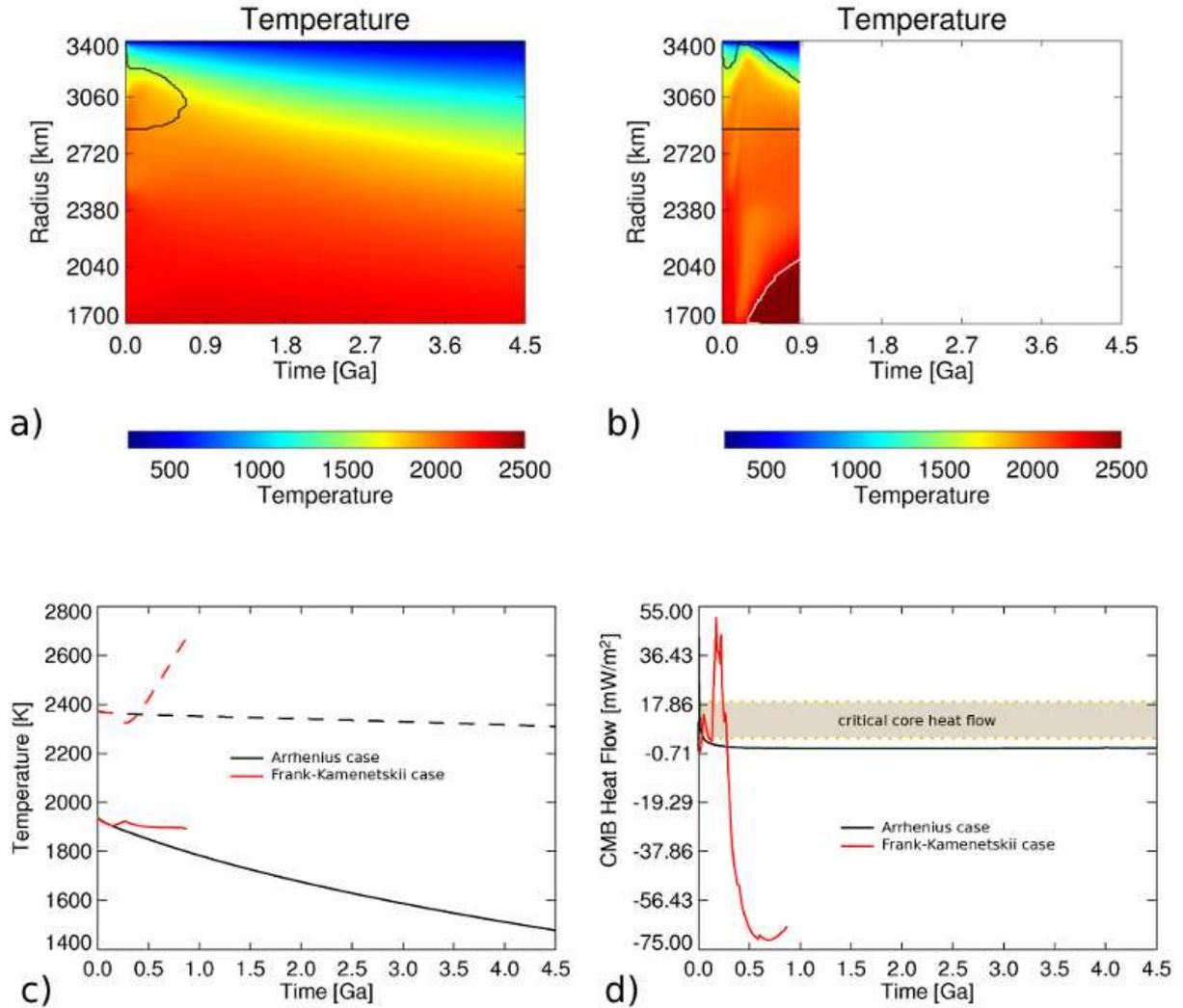}
    \caption{Thermochemical evolution with magma ocean overturn. a) Evolution of the laterally averaged temperature for the Arrhenius and b) F-K case. Black and white contours refers to regions 
    where the temperature lies above the solidus and liquidus, respectively. c) Evolution of the mean mantle temperature (full lines) and CMB temperature (dashed lines). d) Evolution of the CMB 
    heat flow with the interval of critical core heat flow after \citet{nimmo00} marked in gray. In both simulations, we used a reference viscosity of $10^{22}$ Pa s, a plastic rheology with a surface yield stress of $10^8$ Pa, and a yield stress gradient of $160$ Pa/m, an initial temperature profile from \citet{elkinstanton05b}, and considered the entire amount of heat sources 
    to be located in the uppermost $50$ km of the mantle.} 
    \label{fig_magmaocean_evolution}
 \end{center}
\end{figure}

In the third case, the density profile considering phase transitions is even more unfavorable for whole-mantle overturn than the previous one. Similarly to the second simulation, the overturn only involves the mantle below the stagnant lid. Upwelling material from the lower part of the mantle is blocked by lighter cumulates residing in the upper part. The garnet layer located in the upper mantle changes its density only slightly when undergoing the phase transition at greater depths. \add{As a consequence, and in contrast to the results neglecting the phase transitions, the garnet layer does not sink to the core-mantle boundary but remains in the upper part of the lower mantle just below the phase transition. The smallest density of the lower mantle material, after this material rises and undergoes the phase transition, is about $3450$ kg/m$^3$, thus still higher than the upper mantle densities below the stagnant lid (see Figure \ref{fig_magmaocean_profile}). Some material from the lower part of the mantle moves above the garnet layer whose density is about $3660$ kg/m$^3$. However, only a small amount of material can be exchanged between the upper and lower mantle as the garnet layer, which remains at approximately the same depth, acts against advective transport.} Exchange can only take place during the overturn and ceases as soon as a stable configuration is reached in the lower part of the mantle. The mantle then cools rapidly to a conductive state as shown in Figure \ref{fig_magmaocean_evolution} and melting in the upper mantle only occurs during the first $1000$ Ma.

\subsection{Arrhenius Law vs. Frank-Kamenetskii Approximation}\label{subsect_viscosity}

In the following models we only use the third chemical density profile discussed in the previous section, and investigate the effects of the rheological law. It is important to note that the results presented in previous studies \citep[e.g.,][]{elkinstanton05a, debaille09} are strongly influenced by the rheological parameters used. For example, the use of a low activation energy and high surface temperature, or of a small viscosity contrast all contribute to considerably reduce global viscosity contrasts. In \citet{elkinstanton05b}, the use of an activation energy of only $100$ kJ/mol and a surface temperature above $1000$ K results in a viscosity contrast of about three orders of magnitude, which shifts the convection regime from stagnant- to sluggish-lid, in which also the surface layer becomes mobile. In contrast to our findings, the densest material located at the surface overturns because of the mobilization of the lithosphere.\add{ More recently, \citet{debaille09} used a higher viscosity contrast ($\gamma=\ln(10^5)$) and showed that, when using a yielding mechanism, the lithosphere undergoes plastic deformation and whole-mantle overturn can take place as in \citet{elkinstanton05b}.} However, the use of the F-K approximation is highly debated, especially when considering plastic yielding as it can alter the viscosity profile in the lithosphere, leading  to erroneous results \citep{noack13}.

\begin{figure}[!ht]
 \begin{center}
    \includegraphics[width=\textwidth,angle=0]{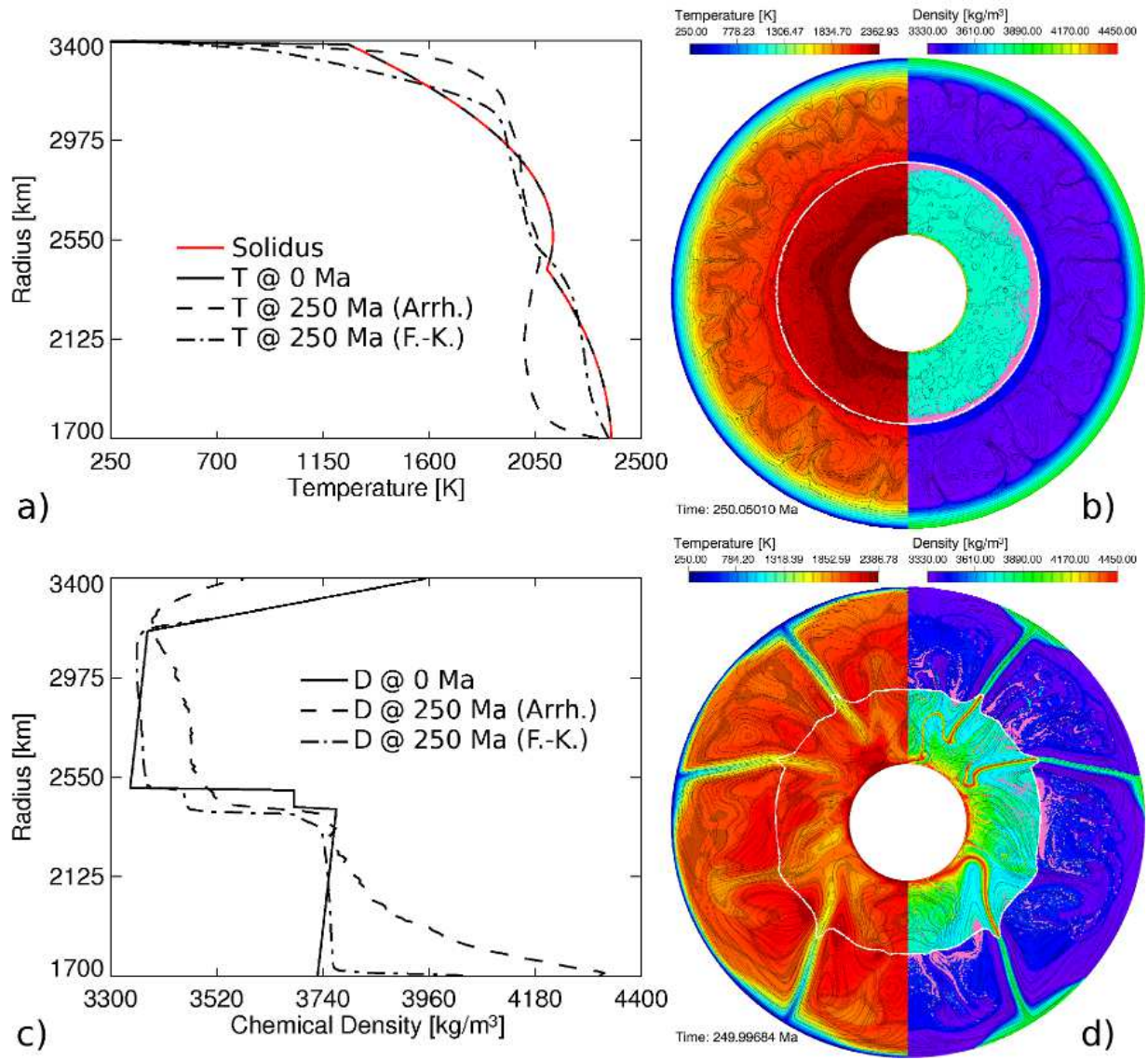}
    \caption{Effect of the rheological law. A case using the Arrhenius law for the viscosity calculation is compared to a case where the Frank-Kamenetskii approximation has been used. a) 
    Temperature and c) density profiles for the Arrhenius  (dotted dashed line) and F-K (dashed line) case at $250$ M. Temperature and chemical density distribution for b) Arrhenius and d) F-K case. While in the Arrhenius case the overturn takes place below the stagnant lid, in the F-K case a whole-mantle overturn occurs. The garnet layer is shown in pink. An animated sequence of both simulations is shown as supplementary material.} \label{fig_viscosity}
 \end{center}
\end{figure}

Indeed, when we use the Arrhenius law with an activation energy of $300$ kJ/mol and a surface temperature of $250$ K, the lithosphere remains rigid and does not experience plastic deformation.\add{ 
If instead we use the F-K approximation with a parameter $\gamma=\ln(10^5)$, the upper dense layer overturn. In this case, the less steep viscosity gradient implied by the F-K approximation
allows for larger deformations at shallow depths than those observed using the Arrhenius law, with the consequence that the yield stress can be locally exceeded.} 
\add{Nevertheless, it should be also noted that this behavior is valid for a particular choice of the yield stress. A smaller value of the latter would likely allow for the mobilization of the 
surface even when using the Arrhenius law \citep{tosi13}.}
%high surface temperatures, which reduce the viscosity contrast considerably. Therefore the convection is in a sluggish regime including also the surface layer. However, assuming that a initial steam atmosphere will be rapidly lost during or after the magma ocean fractional crystallization, we use a low surface temperature, which then results in a stagnant lid at the top of the mantle.
%In a fist test we model the viscosity by using the Arrhenius law with a viscosity cut when the viscosity reaches the value $10^{32}$ Pa s. A second run uses the FK approximation setting the reference temperature to $T_{ref} = 1600$ K and applying a viscosity contrast of $10^5$ as used in previous studies \citep[e.g.,][]{debaille09}. In both cases we use a depth dependent yield stress to allow for the stagnant lid to break.
In Figure \ref{fig_viscosity} we plot the temperature and chemical density profiles from both runs after $250$ Ma. In the Arrhenius case, the overturn occurs below the stagnant lid, while in the F-K case also the uppermost dense layers sink to the CMB. This can be seen both in the chemical density profile where the densest material is at the CMB, as well as in the temperature profile, where the cold surface material carried along by the chemically dense upper layer lies in the lower part of the mantle. The garnet layer rises to shallower depths only if surface mobilization takes place. In the absence of dense cumulates sinking from the top of the mantle to produce a strong return flow, this layer remains approximately at its original depth (compare Figure \ref{fig_viscosity}b and \ref{fig_viscosity}d; for a more detailed time evolution of the garnet layer see online supplementary material). Dynamically, this is quite the opposite as what observed in the models presented by \citet{debaille09}, where this garnet-rich layer shifts to the upper mantle during the overturn phase, being pushed by the sinking uppermost dense material. 
%Since in our simulation using the Arrhenius viscosity law the densest material is trapped in the stagnant lid, it will not affect in any way the overturn dynamics.

\add{The presented F-K case can however be used to study the consequences of the global mantle overturn with surface mobilization on the subsequent thermal evolution -- in case different mechanisms not considered here would lead to an early surface mobilization.} During the overturn, the uppermost dense material sinks and accumulates at the CMB. The entire amount  of heat sources, initially enriched in the uppermost $50$ km, resides, after the overturn, above the CMB. Due to the stable density gradient established after the overturn, thermal convection is difficult to maintain. Although the upper mantle cools, the lower part of the mantle is strongly affected by the stable density gradient and by the enrichment of heat producing elements above the CMB. This results in a strong increase of the temperature to values that even exceed the liquidus (Figure \ref{fig_magmaocean_evolution}b). Although our model does not include the effects of partial melting, we argue that even if melting were considered, the depth where this occurs would be below the density inversion depth \citep{ohtani95} where the melt, due to its compressibility, is negatively buoyant. Therefore it would either remain in suspension with the silicate matrix or sink to the CMB.

%%%%%%%%%%%%%%%%%%%%%%%%%%%%%%%%%%%%%%%%%%%%%%%%%%%%%%%%%%%%%%%%%%%%%%%%%%%%%%%%%%%%%%%%%%%%%%%
%%    			Discussion
%%%%%%%%%%%%%%%%%%%%%%%%%%%%%%%%%%%%%%%%%%%%%%%%%%%%%%%%%%%%%%%%%%%%%%%%%%%%%%%%%%%%%%%%%%%%%%%
\section{Discussion and Conclusions}
\label{Discussion}

We have investigated the consequences of an unstable density profile resulting from fractional crystallization of a global magma ocean on the thermal evolution of Mars. We have found that considering a \add{strongly} temperature-dependent rheology \add{with a high activation energy}, plastic yielding \add{with a surface yield stress of $10^8$ Pa}, and a surface temperature close to present-day value as suggested by recent modelling efforts \citep{lebrun13,erkaev13}, a global mantle overturn with surface mobilization is unlikely. \add{Note, however, that even a high surface temperature of $900^{\circ}$C, which can be maintained for longer than $10$ Ma according to \citet{elkinstanton05b} and using an activation energy of $300$ kJ/mol, as expected for an olivine-dominated rheology, does not show surface mobilization.} This contrasts with the findings of \citet{elkinstanton03, elkinstanton05b, debaille09} who used a density profile that did not account for phase transitions, and a simplified viscosity approximation. Mars most likely has been operating in a stagnant lid regime over its entire evolution. \add{Our simulations show that, under the hypothesis of magma ocean fractional crystallization,} the densest cumulates remained trapped in the stagnant lid and only an overturn below it took place. The underlying mantle evolved into two regions separated by the garnet layer. Convection and partial melting ceased during the first $1$ Ga since the majority of the heat sources are also trapped in the stagnant lid. Although stable chemical reservoirs form early in the evolution -- as required by the isotopic data of the martian meteorites -- the described scenario has several drawbacks that appear to be inconsistent with our current knowledge of the planet. The strong chemical gradient implied by fractional crystallization \citep{elkinstanton05b, debaille09} surppresses convection and partial melting. On the one hand, it is difficult to trace any reservoir in the lower part of the mantle. On the other hand, the rapid decline of convection and partial melting is at odds with the expected long-lived volcanic activity \citep[e.g.,][]{neukum04}. Furthermore, a dense primordial lid is difficult to reconcile with the low crustal density estimated for the southern hemisphere \citep{pauer08}. Such scenario is also not consistent with the formation of nakhlites as proposed in \citet{debaille09}, since the garnet layer can not rise to shallow depths and the material most likely will not experience high melt fractions for garnet segregation to occur. 

Although we argue that surface mobilization at the end of the magma ocean phase is unlikely, it remains a possibility that can not be ruled out. For instance, a dense atmosphere and a high surface temperature close to or higher than the solidus of silicates may allow surface material to be mobilized -- this requires the overturn of the magma ocean within a time shorter than $10$ Ma \citep{elkinstanton08}. \add{Isotopic analyses of the SNC meteorites suggest, however, source reservoirs differentiation between $10$ Ma \citep{shih99, foley05} and $100$ Ma \citep{blichert99, debaille09}.} Similarly, external processes such as impacts may trigger the instability of the upper dense layer. In that case, the whole mantle overturn leads to a stable density profile characterized by a basal layer highly enriched in radiogenic heat sources. This results in a strong overheating of the lower mantle up to above the liquidus temperature.\add{ It is important to note that heating of this dense layer above the CMB does not destabilize it as suggested for example for the ilmenite-rich layer resulting from the solidification of the lunar magma ocean \citep[e.g.,][]{stegman03}. In the latter case, the difference in density between ilmenite-rich layer and the overlying mantle is assumed to be only $90$ kg/m$^3$ \citep{stegman03,zhang13}, and hence much lower than the overall density contrast
we considered in our models ($\sim 800$ kg/m$^3$ following \citet{elkinstanton05b}).} 
%The temperature difference needed to overcome \add{this large}\del{the} density gradient reached after the overturn is in the range of a few tens of thousand of K. 

The melt generated in this lower part of the mantle is highly enriched in iron, since the densest layer contains iron-rich cumulates. Moreover, this melt is located below the density inversion depth -- the depth below which melt is denser than the surrounding silicate matrix. \add{For Mars, due to the high Fe-O content of the melts, this depth has been estimated to lie at $7$ GPa, corresponding to a depth of about $540$ km \citep{ohtani95}}. Therefore, we speculate that this melt would remain trapped in the lower part of the mantle and, because of the slow cooling of this region, would probably persist until recent times (perhaps until present). The upper mantle instead rapidly cools to a conductive state due to the absence of heat sources. Volcanic activity is expected to end in the first few hundred million years, at odds with the martian volcanic history \citep[e.g.,][]{werner09}.

In both described scenarios, i.e. whole mantle overturn and overturn below the stagnant lid, we obtain the formation of stable chemical reservoirs. Nevertheless, because of the strong chemical gradient assumed by the fractional crystallization model \citep{elkinstanton05b, debaille09}, it is difficult to trace these reservoirs, particularly in the lower part of the mantle. Furthermore, our results can not be reconciled with the subsequent thermo-chemical evolution of Mars. Therefore, we conclude that a purely fractional crystallization most likely did not take place. If a global magma ocean existed, a combination of fractional and equilibrium freezing, allowing for a smaller density constrast, appears more plausible. In fact, in recent study, \citet{tosi13} shows that for a lower density gradient ($27-244$ kg/m$^3$) a chemical signature can be preserved in the lower part of the mantle, while convection homogenizes with time the upper mantle. In this case, the chemical heterogeneities can be traced by material entrainment from the deep reservoirs, while convection in the upper mantle may be responsible for late volcanic activity.  

Another possibility would be that of a local or hemispherical magma ocean. This could be the result of a large impact in the early stage of the planetary evolution that has been for instance suggested to explain the crustal dichotomy \citep[e.g.][]{golabek11,andrews-hanna08,nimmo08,marinova08}. Such a scenario could possibly reduce the compositional gradient if \add{only local parts of the mantle have been crystallized allowing for efficient mixing of mantle cumulates.\add{ In fact, isolated lithospheric oxygen reservoirs, inferred by the analysis of the recently recovered NWA7034 meteorite, also support the scenario of impact driven regional magmatic processes instead of a global magma ocean which would have homogenized these reservoirs \citep{agee13}.} In addition, density variations in the mantle due to partial melting that are associated with density contrasts of about $60$ kg/m$^3$ between primordial and depleted mantle  can explain the early formation of long-lasting reservoirs \citep[]{schott01, ogawa11, plesa13a}. In fact, for a better understanding of the martian thermo-chemical evolution, both the likely density stratification after solidification of a magma ocean and by partial melting need to be considered. }

\section*{Acknowledgments}

\noindent 
\add{We wish to thank Lindy Elkins-Tanton and an anonymous reviewer for their constructive comments, which helped to improve an earlier version of this manuscript.  }
This research has been supported by the Helmholtz Association through the research alliance "Planetary Evolution and Life", by the Deutsche Forschungs Gemeinschaft (grant TO 704/1-1), by the Interuniversity Attraction Poles Programme initiated by the Belgian Science Policy Office through the Planet Topers alliance, and by the High Performance Computing Center Stuttgart (HLRS) through the project \textit{Mantle Thermal and Compositional Simulations (MATHECO)}.

%%%%%%%%%%%%%%%%%%%%%%%%%%%%%%%%%%%%%%%%%%%%%%%%%%%%%%%%%%%%%%%%%%%%%%%%%%%%%%%%%%%%%%%%
%   Begin Bibliography
%%%%%%%%%%%%%%%%%%%%%%%%%%%%%%%%%%%%%%%%%%%%%%%%%%%%%%%%%%%%%%%%%%%%%%%%%%%%%%%%%%%%%%%%
%\bibliographystyle{elsarticle-harv}
%\bibliography{bibl}

%%%%%%%%%%%%%%%%%%%%%%%%%%%%%%%%%%%%%%%%%%%%%%%%%%%%%%%%%%%%%%%%%%%%%%%%%%%%%%%%%%%%%%%%
%   End Bibliography
%%%%%%%%%%%%%%%%%%%%%%%%%%%%%%%%%%%%%%%%%%%%%%%%%%%%%%%%%%%%%%%%%%%%%%%%%%%%%%%%%%%%%%%%

%%%%%%%%%%%%%%%%%%%%%%%%%%%%%%%%%%%%%%%%%%%%%%%%%%%%%%%%%%%%%%%%%%%%%%%%%%%%%%%%%%%%%%%%
%   Figures and Tables
%%%%%%%%%%%%%%%%%%%%%%%%%%%%%%%%%%%%%%%%%%%%%%%%%%%%%%%%%%%%%%%%%%%%%%%%%%%%%%%%%%%%%%%%

\newpage
\begin{longtable}{l|c|c}%[!ht]
\centering
%\begin{tabular}{c|c|c}
Parameter & Symbol & Value\\
\hline
Planet radius $[km]$ & $R_{p}$ & $3400$ \\
Core radius $[km]$ & $R_{c}$ & $1700$ \\
Mantle thickness $[km]$ & $D$ & $1700$ \\
Temperature $[K]$ & $T$    & - \\
Surface temperature $[K]$ & $T_{surf}$	& $250$ \\
Reference temperature $[K]$ & $T_{ref}$    & $1600$ \\
Mantle density $[kg/m^3]$ & $\rho_m$ & $3535$\\
Gravitation $[m/s^2]$ & $g$ & $3.7$ \\
Thermal expansion $[1/K]$ & $\alpha$ & $2.5\cdot 10^{-5}$ \\
Temperature drop across the mantle $[K]$ & $\Delta T$ & $2123.15$ \\
Thermal conductivity $[W/mK]$ & $k$ & $4$ \\
Thermal diffusivity $[m^2/s]$ & $\kappa$ & $10^{-6}$ \\
Core density $[kg/m^3]$ & $\rho_c$ & $7200$ \\
Specific mantle heat capacity $[J/kg K]$ & $c_p$ & $1200$ \\
Specific core heat capacity $[J/kg K]$ & $c_{p,c}$ & $840$ \\
Viscosity $[Pa$ $s]$ & $\eta$ & - \\
Reference viscosity $[Pa$ $s]$ & $\eta_{ref}$ & $10^{22}$ \\
Initial amount of radiogenic heat sources $[pW/kg]$ & $Q$ & $23$ \\
Density contrast across the mantle $[kg/m^3]$ & $\Delta \rho$ & $620$ \\
Composition $[$-$]$ & $C$ & - \\
Dissipation number $[$-$]$ & $Di$ & $0.131$ \\
Activation energy $[kJ/mol]$ & $E$ & $300$ \\
Surface yield stress $[Pa]$  & $\sigma_{0}$ & $10^8$ \\
Yield stress gradient $[Pa/m]$  & $d\sigma_{y}/dz$ & $160$ \\
Viscous dissipation $[W/m^3]$  & $\Phi$ & - \\
Velocity $[m/s]$  & $u$ & - \\
Radial velocity $[m/s]$  & $u_r$ & - \\
Dynamic pressure $[Pa]$  & $p$ & - \\
Time $[s]$  & $t$ & - \\
Thermal Rayleigh number $[$-$]$ & $Ra$ & $3.41\cdot 10^5$ \\
Heat Sources Rayleigh number $[$-$]$ & $Ra_Q$ & $9.43\cdot 10^6$ \\
Chemical Rayleigh number $[$-$]$ & $Ra_C$ & $1.12\cdot10^6$\\
Clapeyron slope $[MPa/K]$ & $\gamma_0$ & $2.9$\\
Phase transition reference depth $[km]$ & $z_0$ & $1000$\\
Phase transition reference temperature $[K]$ & $T_0$ & $2100$\\
Phase transition width $[km]$ & $w$ & $15$\\
%\end{tabular}
\caption{Parameters used in the simulations.}
\label{tab_parameters}
\end{longtable}

\newpage

\newpage

\newpage

\newpage

\newpage

\end{document}